\documentclass[aps,nofootinbib,preprint,tightenlines]{revtex4}

\usepackage{graphicx}
\usepackage{amsmath}
\usepackage{bm} 

\newcommand{\beq}{\begin{eqnarray}}
\newcommand{\eeq}{\end{eqnarray}}
\newcommand{\bqa}{\begin{eqnarray}}
\newcommand{\eqa}{\end{eqnarray}}

\def\mqo2{{\!\!\!}}

\begin{document}


\preprint{HISKP-TH-08/12}
\title{Universal properties and structure of halo nuclei}
\author{David L. Canham}\email{canham@itkp.uni-bonn.de}
\affiliation{Helmholtz-Institut f\"ur Strahlen- und Kernphysik (Theorie),
 Universit\"at Bonn, 53115 Bonn, Germany\\}

\author{H.-W. Hammer}\email{hammer@itkp.uni-bonn.de}
\affiliation{Helmholtz-Institut f\"ur Strahlen- und Kernphysik (Theorie)\\
and Bethe Center for Theoretical Physics,
 Universit\"at Bonn, 53115 Bonn, Germany\\}

\date{\today}


\begin{abstract}
The universal properties and structure
of halo nuclei composed of two neutrons ($2n$) and 
a core are investigated within an effective quantum mechanics framework. 
We construct an effective interaction potential that exploits the
separation of scales in halo nuclei and treat the nucleus as an 
effective three-body system. The uncertainty from higher orders in the 
expansion is quantified through theoretical error bands.
First, we investigate the possibility to observe excited
Efimov states in $2n$ halo nuclei. Based on the experimental data, 
$^{20}$C is the only halo nucleus candidate to possibly have an Efimov 
excited state, with an energy less than 7~keV below the scattering threshold.
Second, we study the structure of $^{20}$C and other $2n$ halo nuclei.
In particular, we calculate their matter form factors, radii, and 
two-neutron opening angles.
\end{abstract}

\maketitle


\section{Introduction}
\label{sec:Intro}

There has been a considerable interest in physical systems with
large scattering lengths recently.
The scattering of two particles with short-range interactions
at sufficiently low energy is determined by their S-wave scattering length
$a$. If $a$ is much larger than the range of
the interaction $r_0$, the system shows universal
properties \cite{Braaten-05}. 
The simplest example is the existence of a shallow 
two-body bound state if $a$ is large and positive, but there are many more, 
including the effects of a limit cycle \cite{Bedaque:1998kg}
and the Efimov effect \cite{Efimov-70} in the three-body system.

The best known example of a nuclear system with a large scattering
length is the two-nucleon ($NN$) system. There are two independent S-wave
scattering lengths that govern the low-energy scattering
of nucleons. 
Both scattering lengths are significantly larger than the range of
the interaction $r_0\sim 1/m_\pi \approx 1.4$ fm, while
the effective ranges are of the same order as $r_0$.
As a consequence, the description of few-nucleon systems in an 
expansion in $r_0/|a|$ is useful. It has successfully been applied 
to various two-, three-, and four-nucleon observables 
(See Refs.~\cite{Bedaque:2002mn,Bedaque:2002yg,
Platter:2004zs} and references therein).

Another type of system that can be described with these techniques are halo nuclei:
a special class of nuclear systems which offer the possibility of exploring universal 
behavior \cite{Riisager-94, Hansen-95, Taniha-96,Zhukov-93, Jensen-04}.  
Halo nuclei consist of a tightly bound core and a \lq\lq halo'' of lightly bound nucleons. They are characterized by 
their large nuclear radius compared to the radius of the core. 
Equivalently, the separation energy of the halo nucleons is small 
compared to the excitation energy of the core. 
This separation of scales allows for the use of 
effective theories, where one can assume the core to be a structureless particle, and treat the nucleus as a few-body system of
the core and the valence nucleons.

The first application of effective field theory methods to halo nuclei was carried out in Refs.~\cite{Bertulani-02,BHvK2}, where the 
neutron-alpha system (``$^5$He'') was considered. More recent studies have focused on the consistent inclusion of the 
Coulomb interation in two-body halo nuclei such as the proton-alpha \cite{BRvK} and alpha-alpha systems \cite{Higa:2008dn}.
Three-body halo nuclei composed of a core and two valence neutrons
are of particular interest due to the possibility of these systems to display the 
Efimov effect \cite{Efimov-70}. Efimov found that in three-body systems of non-relativistic particles, if at 
least two of the three pairs of particles have a large scattering length $|a|$ compared to the 
range $r_0$ of the interaction, there occurs a sequence of three-body bound states whose binding energies are 
spaced geometrically between $\hbar^2 / m r_0^2$ and $\hbar^2 / m a^2$. 
The number of bound states grows to infinity, with an accumulation point at the three-body scattering 
threshold, in the limit $\pm a \rightarrow \infty$. The sequence of three-body bound states have universal 
properties that are independent of the details of the two-body potential at short distances. 
The influence of long-range Coulomb interactions on the geometric 
bound state spectrum 
was recently investigated in a model study \cite{Hammer:2008ra}.

The first experimental evidence for Efimov states in ultracold Cs atoms has recently been found through their signature
in three-body recombination rates \cite{Kraemer-06}. This signature could be unravelled by varying the scattering length $a$
over several orders of magnitude using a Feshbach resonance. For halo nuclei, the interaction strength can not easily be
varied and one has to look for different signatures of the Efimov effect. Since the ground state of a halo nucleus can not be 
uniquely identified
as an Efimov state for fixed $a$, it is customary to look for excited states that have the Efimov character. One can then consider
a halo nucleus to display the Efimov effect if it has at least one excited state with universal properties.

In this paper, we explore the occurance of the Efimov effect and its well known universal 
properties for 2$n$ halo nuclei with a core of spin zero. 
From the earliest studies of halo nuclei, $^{20}$C has been suggested as a good candidate for Efimov states \cite{Fedorov-94}, 
with future theoretical work supporting this prediction \cite{Amorim-96, Mazumdar-00}. 
We critically examine these earlier studies and perform an improved analysis in the framework of an effective theory.
The uncertainties of our leading order analysis are quantified through error bands.
In the second part of this paper, we focus on the structure of 2$n$ halo nuclei. In particular, we 
calculate their matter form factors, radii, and two-neutron opening angles. 
Finally, we end with conclusions and an outlook.

\section{Theoretical framework}
\label{sec:Equations}

For our study, we choose the effective quantum mechanics framework of Refs.~\cite{Platter:2004qn,Platter:2004zs,Lucas-04}
which is equivalent to using field-theoretic language for the problem at hand.
The short-range interactions characteristic of halo nuclei are then described using an effective interaction potential. 
The low-energy behavior of the system can then be reproduced with a level of accuracy proportional to 
powers of the low-momentum scale $M_{low}$ over the high-momentum scale $M_{high}$.
The theory is valid up to a momentum, $M_{high}$, at which the errors 
become of order one.  
For example, the two-body interactions of 
halo nuclei can be characterized by their large scattering lengths, $a\sim 1/M_{low}$, and their range, $r_0 \sim  1/M_{high}$.  
Such systems need to leading order
one coupling parameter, $C_0$, for each pair interaction tuned to reproduce the scattering length.
The range of the interaction enters at next-to-leading order.
For $a > 0$, there is a two-body bound state, and the binding energy can then be found through the universal formula:
\beq
B_2 = {\hbar^2 \over 2\mu a^2}+\ldots\,, 
\label{Efroma}
\eeq 
where $\mu$ is the reduced mass. The dots indicate corrections of the order $\sqrt{2\mu B_2}/M_{high}\sim r_0/a$.

For the large separation of scales involved in halo nuclei, zero-range interactions 
can be used in constructing the effective interaction potential.  
This leads to a separable potential made up of contact interactions 
in a momentum expansion. The 2-body S-wave potential to leading order is 
\beq
\langle \vec{p} \mid V_{eff} \mid \vec{p}' \rangle = C_0\, g(p) g(p')+\ldots.
\label{SwaveVeff}
\eeq
where the dots indicate higher order momentum dependent interactions which we will neglect. In a future study, the errors could 
be systematically reduced by including the effective range correction to Eq.~(\ref{SwaveVeff}).
$g(p)$ is the regulator function (sometimes called the form factor) of the theory.  
Of course, the low-energy observables must be independent of the regularization scheme, and one can choose the
scheme most convenient for calculations. We use a momentum cutoff scheme, multiplying the coupling parameter 
with a Gaussian regulator function
\beq
g(p) = \exp\left(-{p^2 \over \Lambda^2}\right), 
\label{reg}
\eeq
where $\Lambda$ is the cutoff parameter.
This regulator function obviously suppresses the contributions of momenta $p,p' \gg \Lambda$, where the 
effective potential would break down and no longer be valid.  A natural choice for the value of $\Lambda$ is therefore
$\Lambda \sim M_{high}$, but observables are independent of $\Lambda$ after renormalization.

This potential is then used in the solution of the three-body Faddeev equations in terms of the spectator functions
$F_i(q)$, which represent the dynamics of the core ($i=c$) and the halo neutron ($i=n$). To find 
the bound state of a halo nucleus composed of two valence neutrons and a core with spin zero, the resulting 
coupled integral equations are simply a generalization of the three-boson equation\footnote{In fact, 
the equations are the same for any bound three-body system of two identical particles and a core with spin zero, 
which interact through the pair-wise zero-range potentials given in Eq.~(\ref{SwaveVeff}).} 
(see \cite{Platter:2004qn} and references within).  We use units such that $\hbar =c = 1$ and the nucleon mass $m=1$:  
\bqa
F_n(q) & =& \ {1 \over 2} \int_0^\infty dq'q'^2 \int_{-1}^1 dx  \bigg[ g\left(\tilde{\pi}(q,q')\right) g\left(\tilde{\pi}(q',q)\right) 
G^n_0\left(\tilde{\pi}(q',q),q';B_3\right)  t_n(q';B_3) F_n(q') 
\nonumber\\
& & + \ g\left(\tilde{\pi}_1(q,q')\right) g\left(\tilde{\pi}_2(q,q')\right) 
G^c_0\left(\tilde{\pi}_2(q,q'),q';B_3\right)  t_c(q';B_3) F_c(q') \bigg],
\label{Fn}
\\
F_c(q) & =& \ \int_0^\infty dq'q'^2 \int_{-1}^1 dx\, g\left(\tilde{\pi}_1(q',q)\right) g\left(\tilde{\pi}_2(q',q)\right) 
G^n_0\left(\tilde{\pi}_1(q',q),q';B_3\right)  t_n(q';B_3) F_n(q'),\nonumber\\
\label{Fc}
\eqa
where the shifted momenta $\tilde{\pi}$, $\tilde{\pi}_1$, and $\tilde{\pi}_2$ are given by:
\bqa
\tilde{\pi}(q,q') = \sqrt{\left({1 \over A+1}\right)^2 q^2 + q'^2 + {2 \over A+1}\,qq'x},
\label{pitilde}
\\
\tilde{\pi}_1(q,q') = \sqrt{\left({A \over A+1}\right)^2 q^2 + q'^2 + {2A \over A+1}qq'x},
\label{pitilde1}
\eqa
and
\beq
\tilde{\pi}_2(q,q') = \sqrt{q^2 + {1 \over 4}q'^2 + qq'x}\,.
\label{pitilde2}
\eeq
In the above equations, $B_3>0$ is the three-body bound state energy and $A$ 
is the number of nucleons in the core.
The free three-body propagators for a spectator neutron $G_0^n$ and a spectator core $G_0^c$ are:
\bqa
G_0^n(p,q;B_3) &=& \left[B_3 + {A+1 \over 2A}p^2 + {A+2 \over 2(A+1)}q^2\right]^{-1},
\label{propn}\\
G_0^{c}(p,q;B_3) &=& \left[B_3 + p^2 + {A+2 \over 4A}q^2\right]^{-1}\,.
\label{propc}
\eqa
The effects of the interactions are contained in the two-body T-matrices which are obtained by solving the 
Lippmann-Schwinger equation for the neutron-neutron and neutron-core interaction with the
effective potential, Eq.~(\ref{SwaveVeff}). In the kinematics of Eqs.~(\ref{Fn}, \ref{Fc}), we have:
\beq
t_n (q';B_3) & = & {1 \over \pi} {A+1 \over A} \left[ -{1 \over a_{nc}} \exp\left({2 / (a_{nc}^2\Lambda^2)}\right) 
\textnormal{erfc}\left({\sqrt{2} /(|a_{nc}| \Lambda)}\right) \right. \nonumber\\
\nonumber\\
& & \left. + \sqrt{\tilde{E}_n(q';B_3)} \exp\left({2\tilde{E}_n(q';B_3)/ \Lambda^2}\right) 
\textnormal{erfc}\left({\sqrt{2\tilde{E}_n(q';B_3)} /\Lambda}\right) \right]^{-1},
\label{tn}
\eeq
with
\beq
\tilde{E}_n(q';B_3) = {2A \over A+1}\left(B_3 + {A+2 \over 2(A+1)}q'^2 \right),
\label{En}
\eeq
and
\beq
t_c (q';B_3) & = & {2 \over \pi} \left[ -{1 \over a_{nn}} \exp\left({2 /(a_{nn}^2\Lambda^2)}\right) 
\textnormal{erfc}\left({\sqrt{2} /(| a_{nn}| \Lambda)}\right) \right. \nonumber\\
\nonumber\\
& & \left. + \sqrt{\tilde{E}_c(q';B_3)} \exp\left({2\tilde{E}_c(q';B_3) /\Lambda^2}\right) 
\textnormal{erfc}\left({\sqrt{2\tilde{E}_c(q';B_3)}/\Lambda}\right) \right]^{-1},
\label{tc}
\eeq
with
\beq
\tilde{E}_c(q';E) = B_3 + {A+2 \over 4A}q'^2.
\label{Ec}
\eeq
where the index $n$, $c$ indicates the spectator particle. 
In Eqs.~(\ref{tn},~\ref{tc}), $\textnormal{erfc}(x)=1-(2/\sqrt{\pi})\int_0^x \exp(-t^2) dt$ denotes 
the complementary error function, which will go quickly to $1$ for $x \ll 1$.
If the cutoff is chosen large compared to all momentum scales involved in the problem: $\Lambda \gg 1/|a|, \sqrt{ \tilde{E}}$, 
these T-matrices reproduce the usual effective range expansion at leading order.

A two-body bound state appears as a simple pole in the two-body t-matrix at an energy $E = -B_2$.  
In our renormalization of the coupling parameter, we have tuned $C_0(\Lambda)$ so that it reproduces the 
scattering length needed to produce this two-body binding energy to leading order, as given in Eq.~(\ref{Efroma}): 
\beq
{1 \over C_0} & = & 2\pi^2 2\mu \left[{1 \over a}  \exp\left({2 \over a^2\Lambda^2}\right) \textnormal{erfc}\left({\sqrt{2} \over |a| \Lambda}\right) - {\Lambda \over 2}\sqrt{{2 \over \pi}} \right].
\label{lambda0toa}
\eeq
This renormalization works equally well for the case of virtual states, $a < 0$. 
Also, for $\Lambda \gg 1/|a|$, the exp and erfc functions both quickly approach $1$, and the resulting 
relations are analogous to Ref.~\cite{Braaten-05}. 

For most halo nuclei, the S-wave scattering length is not as well known as the two-body bound (virtual) state energy. 
Therefore, we will generally use the two-body energies, $E_{nn}$ and $E_{nc}$, as input parameters, 
from which we can calculate the scattering length to leading order through Eq.~(\ref{Efroma}), 
$1/a_{ni} = \pm \sqrt{2\mu_{ni} E_{ni}}$, where the $+$ refers to a bound state and the $-$ to a virtual state, 
$i = n$ or $c$, and $\mu_{ni}$ is the corresponding reduced mass.
The difference between the two-body energy and the S-wave scattering length is higher order in the expansion
in $M_{low}/M_{high}\sim r_0/|a|$.

\section{Universal properties}
\label{sec:univprop}

\subsection{The Efimov effect in 2$n$ halo nulcei}
\label{sec:Efimoveffect}

The three-body binding energies are given by the values of $B_3$ for which the coupled integral equations, 
Eqs.~(\ref{Fn}, \ref{Fc}), have a nontrivial solution. In principle, Eqs.~(\ref{Fn}, \ref{Fc}), should also 
include a three-body force term which is required for proper renormalization. However, due to the limit
cycle behavior of this three-body force, it is always possible to choose a cutoff where the three-body force
vanishes. As a consequence, we can simply drop the three-body force and work with a finite cutoff $\Lambda$ 
as in Ref.~\cite{Platter:2004qn}.
Tuning this cutoff to reproduce a given three-body observable, we can predict other low-energy observables
by using the same cutoff \cite{Braaten-05, Kharchenko-73}.  
The cutoff is inversely related to the interaction radius (see \cite{Efimov-90} and references within), thus taking the cutoff to 
infinity is equivalent to taking the range of our potential to zero. It is at this limit 
that the Thomas collapse \cite{Thomas-35} will occur and the energy of the three-body ground state will diverge.  

The Thomas collapse is closely related to the Efimov effect in that the deepest three-body bound states of the 
Thomas collapse can be identified with the deepest Efimov states \cite{Adhikari-88}.  The sequence of three-body Efimov 
states can be found from our integral equations with sufficiently large scattering lengths by finding the 
spectrum of binding energies for a fixed cutoff.  By increasing the cutoff 
new three-body bound states appear in the spectrum at critical values of $\Lambda$, which are geometrically separated. Accordingly, 
the energies of the deeper bound states increase in magnitude. The Thomas effect is seen by the 
divergence of the deepest bound state energy for $\Lambda \rightarrow \infty$.  
However, the states below the natural cutoff, $1/(mr_0^2)$, are physically irrelevant. They are outside
the range of validity of the effective theory and can be ignored.

\begin{figure}[t]
	\centering
		\includegraphics*[width=10cm]{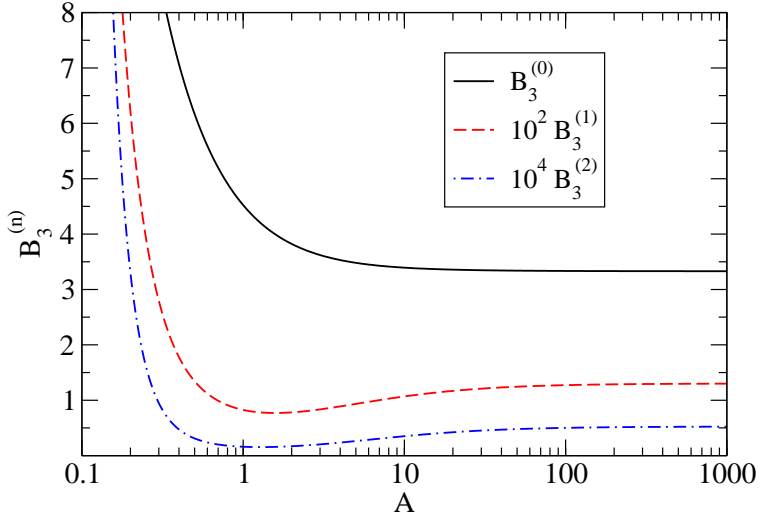}
	\caption{Spectrum of three-body bound states, when two of the particles have equal mass $m_1 = m_2$, as a function of the 
	mass ratio, $A = m_3/m_1$, in the resonant limit $a \rightarrow \pm \infty$.  The cutoff parameter 
	was fixed at a value of $\Lambda = 10.0$ (the units of $\Lambda$ and $B_3^{(n)}$ are arbitrary, for details see text).}
	\label{fig:BvsA}
\end{figure}

Conversely, the sequence of three-body Efimov states have universal properties that are insensitive to the 
details of the two-body potential at short distances, and hence independent of the cutoff.  One such property is 
that for the resonant limit, $a \rightarrow \pm \infty$, at which there are infinitely many arbitrarily-shallow 
three-body bound states, the ratio of the binding energies of neighboring bound states approaches a universal 
factor, $\lambda_0$, as the threshold is approached:
\beq
B_3^{(n)}/B_3^{(n+1)} \longrightarrow ({\lambda_0})^2, \ \textnormal{as} \ n \rightarrow +\infty \ \textnormal{with} \ a = \pm \infty.
\label{Eratio}
\eeq
This universal scaling factor $\lambda_0$ depends on the masses of the particles. In our case, the masses of the two neutrons are equal, 
$m_1 = m_2$, and the core mass  $A = m_3/m_1$.
The values of $B_3^{(n)}$ for $n = 1,2,3$ as a function of the core mass $A$ 
are shown in Fig.~\ref{fig:BvsA} for a finite value of $\Lambda=10$.  
Note that because we have taken $a \rightarrow \pm \infty$, $\Lambda$ is given in units of an arbitrary momentum scale $\kappa$. All
energies are then given in units of $\kappa^2$.
One interesting feature is the appearance of a minimum in the binding energy. 
The absence of this minimum in the $B_3^{(0)}$ curve is due to the fact that the magnitude of $B_3^{(0)}$ is near the order of 
magnitude of $\Lambda$, and details of the regularization scheme become important.  Also, we see that 
the binding energy quickly reaches an asymptotic value for very large $A$, and diverges for $A \rightarrow 0$. 
The dependence of the discrete scaling factor, ${\lambda_0}$, on the core mass $A$ is well known (see, e.g.
the review~\cite{Braaten-05}, Fig.~52). We have checked our code by reproducing these results. In particular,
the discrete scaling factor is largest for all equal masses, where it has the same value $\lambda_0 \approx 22.7$ 
as for three identical bosons. In the very large core limit, $A \gg 1$, the discrete scaling factor approaches $15.7$.  
In the vanishing core limit, $A \ll 1$, it approaches $1$ as all three-body binding energies diverge.  

\subsection{Possibility of Efimov excited states in 2$n$ halo nuclei}
\label{sec:KnKa}

Our main aim in this section is to assess which halo nuclei have the possibility of possessing an 
excited Efimov state. The ground state energy and the two-body energies can not be predicted by our theory and
are taken from experiment. In other words, 
we would generally like to know what the values of the two-body energies must be, or correspondingly 
how large the scattering lengths must be, in order to produce an excited Efimov state, 
knowing the ground state binding energy. 

To this end, we construct the parametric region defined by the ratios, $\sqrt{E_{nc}/B_3^{(n)}}$ versus $\sqrt{E_{nn}/B_3^{(n)}}$. 
The boundary curves representing the existence of an excited Efimov state for various values of the core mass 
are shown in Fig. \ref{fig:KnKaG}. An analogous study was carried out
in Ref.~\cite{Amorim-96} (see below).
All points which lie within the boundary curve have at least one excited 
Efimov state above the state with energy $B_3^{(n)}$, while points outside the curve have no excited states above this state. 
\begin{figure}[!b]
	\centerline{\includegraphics*[width=10cm]{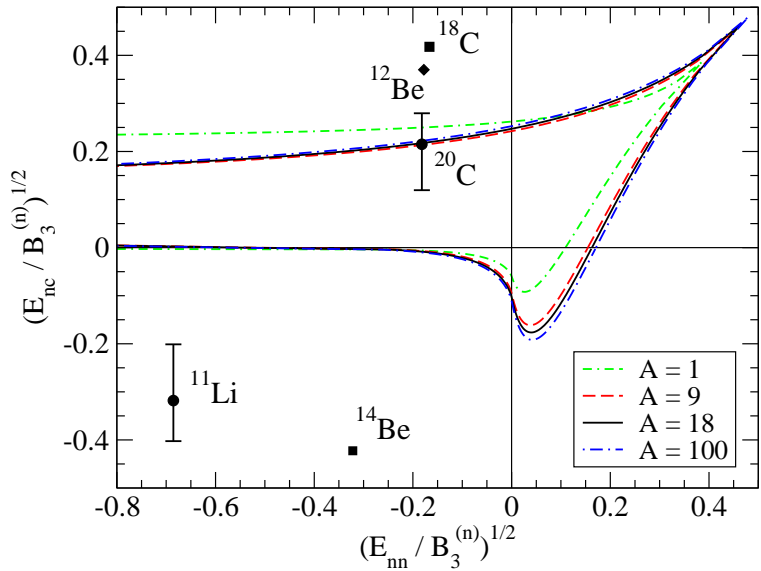}} 
	\caption{Boundary curves in the $\sqrt{E_{nc}/B_3^{(n)}}$ vs. $\sqrt{E_{nn}/B_3^{(n)}}$ plane, where the binding energy 
	of the excited Efimov state $B_3^{(n+1)}$ is exactly at threshold. Negative values on the axes correspond to virtual two-body 
	states. Boundary curves shown for various core masses $A=$1, 9, 18, and 100. 
	Experimental data shown for $^{20}$C, $^{18}$C, $^{11}$Li, $^{12}$Be, and $^{14}$Be are taken from Ref.~\cite{TUNL}.}
	\label{fig:KnKaG}
\end{figure}
The curve itself is built up of the points for which the $B_3^{(n+1)}$ binding energy is equal to the scattering 
threshold; i.e. $B_3^{(n+1)} = E_{ni}$ for $E_{nn}$ or $E_{nc}$ bound, where $E_{ni}$ is the larger of $E_{nn}$ and $E_{nc}$, 
and $B_3^{(n+1)} = 0$ for $E_{nn}$ and $E_{nc}$ virtual. The boundary curves in Fig. \ref{fig:KnKaG} were found with $n = 1$
in order to minimize the regulator effects. Due to the 
scaling symmetry of the sequence of three-body bound states, the $n$th state can always be interpreted 
as the ground state and the ($n+1$)th state as the first excited state.\footnote{Because of the regulator effects, 
the curves found with $n = 0$ are slightly different from the curves in Fig.~\ref{fig:KnKaG}.
The curves are practically the same for 
larger values of $n$, as the numerical values of $B_3^{(n)}$ are much smaller than the cutoff $\Lambda$ for $n > 0$.} 

Here it is of interest to note that these results differ from the results found by Amorim, et al. in an 
analogous study  \cite{Amorim-96} using a hard momentum cutoff rather than the Gaussian regulator scheme. 
We found that the results agree almost exactly for a core mass equal to the nucleon mass, $A = 1$, but differ significantly 
for all other values of the core mass. While the qualitative conclusions on the likelihood of Efimov states 
in 2$n$ halo nuclei are the same, the quantitative results are different. In fact, in doing the numerical calculations 
with a hard cutoff ourselves, we found results that match those presented here.

\begin{figure}[t]
	\centerline{\includegraphics*[width=10cm]{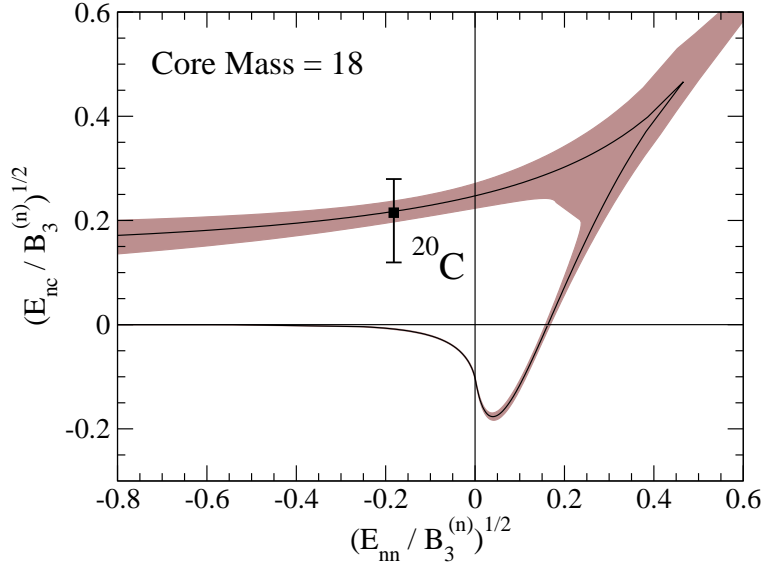}}
	\caption{Boundary curve in the $\sqrt{E_{nc}/B_3^{(n)}}$ vs. $\sqrt{E_{nn}/B_3^{(n)}}$ plane with leading order error bands. 
	Boundary curve shown for a core mass of $A = 18$ with the experimental data for $^{20}$C from Ref.~\cite{TUNL}.}
	\label{fig:KnKaerrors}
\end{figure}

These results represent the leading order calculations with the effective potential described in Eq.~(\ref{SwaveVeff}). 
The theoretical uncertainty in calculating the binding energy is of the order $\approx r_0/a$, 
where $r_0$ is the effective range of the 
potential, and $a$ is the scattering length. As stated before, the cutoff parameter is related to the inverse of the 
range of the potential, such that we can approximate $r_0 \approx 1/\Lambda$. However, it is important in this error estimate 
that the $\Lambda$ used comes from the result with $n=0$, the true ground state, rather than $n=1$. This corresponds to 
taking the \lq\lq natural'' value for the cutoff $\Lambda$. We therefore estimate 
the leading fractional error of our boundary curves as $\approx \sqrt{2\mu_{ni}E_{ni}}/\Lambda$, for $i = n$ and $c$ 
respectively. 
The resulting boundary curve including leading order error bands, using the case of core mass $A = 18$ is shown in 
Fig. \ref{fig:KnKaerrors}. This graph is a good representation of the error bands for other core masses. 
The uncertainty of our leading order calculation becomes large for values of $\sqrt{E_{nn}/B_3^{(n)} + E_{nc}/B_3^{(n)}}$ near 1 and greater. 
At this point  the low-energy observables approach the order of magnitude of $\Lambda$, where the effective potential is no longer 
a good description of the three-body system. 

Now we discuss the implications of Figs.~\ref{fig:KnKaG} and \ref{fig:KnKaerrors} for the existence of excited Efimov states in 
halo nuclei in more detail. The four quadrants of the parametric plane in these figures correspond to the 
four different types of 
three-body halo nuclei, determined by the different types of two-body subsystems. The upper-right quadrant 
corresponds to both the $n$-$n$ and the $n$-$c$ subsystems being bound, and is accordingly called {\it All Bound}. 
The lower-right quadrant is that in which the $n$-$n$ subsystem is bound, but the $n$-$c$ subsystem is unbound, and 
receives the name {\it Tango} \cite{Robicheaux-99}. Of course, because we are concerned with 2$n$ halo nuclei, 
where the $n$ particle is truely 
a neutron, these two quadrants are not of much interest in the present study. The upper-left quadrant corresponds to 
the unbound $n$-$n$ subsystem with a bound $n$-$c$ subsystem, for which we use the name {\it Samba} as recommended in 
\cite{Yamashita-Samba}. The final quadrant corresponds to the three-body systems for which none of the two-body 
subsystems is bound. This system is referred to as a {\it Borromean} system. 

We can now use our plot to analyze the likelihood of the Efimov effect for the different types of three-body systems, 
with a focus on 2$n$ halo nuclei. 
As one would expect, the {\it Borromean} systems offer the smallest chance of having an excited Efimov state, 
as the two-body energy would have to be very small, or accordingly the scattering length very large, to produce 
even one excited state. However, this can be achieved in ultra-cold atoms, as the presence of so called 
{\it Feshbach resonances} allows one to tune the two-body scattering length to a very large value 
\cite{Tiesinga-93}. Interestingly, the {\it Samba} systems have the largest region supporting 
the occurance of excited Efimov states. As long as the $n$-$c$ scattering length is large enough, there can be a 
large variation in the $n$-$n$ scattering length that would still allow for the Efimov effect. This agrees with the 
findings of Efimov \cite{Efimov-70}, that as long as 2 of the 3 two-body interactions have a large scattering length, 
the sequence of three-body binding energies can occur. 

Looking at possible halo nuclei candidates, we have plotted the positions of $^{20}$C, $^{18}$C, $^{11}$Li, $^{12}$Be, 
and $^{14}$Be in Fig.~\ref{fig:KnKaG}, using the experimental values of the "Nuclear Data Evaluation Project" 
of TUNL \cite{TUNL} for the $n$-$c$ and three-body ground state energy data, and the standard value 
of the $n$-$n$ scattering length, $a_{nn} = (-18.7 \pm 0.6)$ fm \cite{Gonzales-99} to calculate 
the $n$-$n$ two-body energy according to Eq.~(\ref{Efroma}). 
The only halo candidate that has any possibility of an excited Efimov state is 
$^{20}$C, due to the large uncertainty in the $n$-$^{18}$C bound state energy. We will return to this nucleus shortly. 
The positions and relatively small experimental errors in the other halo nuclei data rule out the chance of 
finding excited Efimov states in these nuclei. Other halo nuclei candidates which exist have 
values of the two-body energies which are too large to even appear on our plot.  

\subsection{Efimov excited state for $^{20}$C}
\label{sec:Carbon20}

\begin{figure}[t]
	\centering
		\includegraphics*[width=10cm]{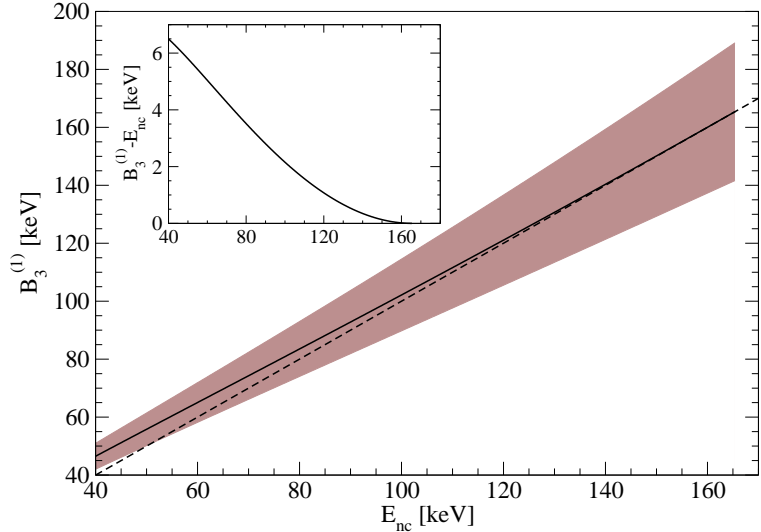}
	\caption{Binding energy of the $^{20}$C excited Efimov state as a function of the $n$-$^{18}$C bound state energy 
	(solid line) with leading order error bands. The dashed line represents the scattering threshold which is given by
        $B_3^{(1)}= E_{nc}$. The inset shows the excited state energy relative to the scattering threshold. }
	\label{fig:20C}
\end{figure}

The central value for the $n$-$^{18}$C bound state energy, $E_{nc} = (162 \pm 112)$ keV \cite{Audi-95}, lies almost 
exactly on the boundary region for $A=18$ in Fig.~\ref{fig:KnKaerrors}. The large error in this value, however, dips well 
into the region where at least one excited Efimov state can occur. The error in the three-body ground state energy of 
$^{20}$C is small compared to $E_{nc}$. Thus, we can calculate the value of the excited state energies as a 
function of $E_{nc}$, using the standard value for $a_{nn}$, and fixing our cutoff to reproduce the experimental 
value of the ground state energy $B_3^{(0)} = 3506.0$ keV \cite{TUNL, Audi-95}. The result is plotted in Fig.~\ref{fig:20C}, 
where the solid line is the excited state energy, and the dashed line represents the scattering threshold.  The inset
graph shows the excited state energy relative to the scattering threshold.
We find only one excited Efimov state, existing when $E_{nc} < 165$ keV. 
For larger values of $E_{nc}$, the $^{20}$C system moves outside of the boundary curve, and 
the excited Efimov state is destroyed. 
The binding energy relative to the scattering threshold is always below 7 keV, a value very small 
in comparison with the ground state energy. Also, the error bands are large compared to the relative energy 
of the exicted state to the scattering threshold, with the lower error band almost always below the scattering 
threshold. 

We have again estimated this error using the theoretical uncertainty of our effective potential. 
In this first order calculation, the uncertainty in binding energies calculated using the two-body effective potential of 
Eq.~(\ref{SwaveVeff}) is $\approx r_0/a$.  Our effective potential, made up of 
contact interactions, will break down for momenta of the order of the pion mass scale. We therefore use the inverse of the pion mass 
$m_{\pi} = 140$ MeV to estimate the effective range $r_0 \approx 1/m_{\pi}$. The uncertainty in the binding energy 
of the excited state is then the quadratic sum
of the uncertainties from the $n$-$n$ and $n$-$^{18}C$ interactions: $\sqrt{2\mu_{nn}E_{nn}/m_{\pi}^2 + 2\mu_{nc} E_{nc}/m_{\pi}^2}$. 
These uncertainties are of the same order of magnitude as those found assuming that the effective 
range is related to the inverse of the cutoff, $r_0 \approx 1/ \Lambda$. 

This result is in overall qualitative agreement with 
the previous studies of Amorim et al.~\cite{Amorim-96} and Mazumdar et al.~\cite{Mazumdar-00}, who also found the 
presence of a very weakly bound excited Efimov state in $^{20}$C for sufficiently small values of $E_{nc}$. 
However, both of these studies have a larger value for the excited state energy, with the Mazumdar 
group also finding a second excited state for $E_{nc} < 100$ keV. Also, the disagreement with the results 
of the Amorim paper mentioned before casts doubts on the quantitative results of \cite{Amorim-96}, 
as a more recent study from the same group \cite{Yamashita-07} suggests better agreement with the results presented here. 
This newer analysis is one of a few recent studies extending the trajectory 
of this excited state into the scattering region to explore the possibility of finding a resonance in the $n$-$^{19}$C 
scattering sector \cite{Yamashita-07, Arora-04, Mazumdar-06}.

\section{Form Factors and Mean Square Radii}
\label{sec:FF}

We are now interested in calculating other low-energy physical properties of three-body halo nuclei, specifically 
the matter density form factors and the mean square radii. The 
information needed to calculate such quantities is held in the wave functions of the known bound states. 
In the previous sections, we only required the Faddeev spectator functions $F_n$ and $F_c$. However, 
the full three-body wave function can be reconstructed from the Faddeev spectator functions (see Appendix \ref{sec:EigtoWF} for 
details).
Once the three-body wave function is known, the corresponding one- and two-body matter density form factors can be computed. 
Finally, the mean square distances between two of the three particles as well as the mean square 
distance of one of the particles from the center of mass can be extracted from the proper form factor. 

\subsection{One- and two-body matter density form factors}
\label{sec:WFtoFF}

The three-body wave functions found in Appendix \ref{sec:EigtoWF} can be used to calculate other low-energy 
properties of the three-body bound state. With the Jacobi momentum states it is straight forward
to calculate the Fourier transform of the one- and two-body matter densities with respect to the momentum 
transfer squared. 
These are defined as the one- and two-body matter density form factors, ${\mathcal F}_{i}(k^2)$ and 
${\mathcal F}_{ni}(k^2)$ respectively, where $i = n,c$. In the wave functions, the $\vec{p}$ Jacobi 
momentum describes the relative momentum between 
the two particles in the chosen two-body subsystem, while $\vec{q}$ describes the momentum of the spectator 
particle relative to the center of mass of the two-body subsystem.\footnote{Recall that we use the spectator notation 
for the wave functions, where the index $i$ on $\Psi_i$ refers to the spectator particle.} Therefore, the one-body form factors 
can be obtained as follows:
\beq
{\mathcal F}_i(k^2) = \int d^3p \ d^3q \ \Psi_i(\vec{p}, \vec{q}) \Psi_i(\vec{p},\vec{q}-\vec{k}),
\label{FF1}
\eeq
where $i = n,c$ depending on the desired two-body subsystem. 
The two-body form factors can be solved similarly:
\beq
{\mathcal F}_{nc}(k^2) = \int d^3p \ d^3q \ \Psi_n(\vec{p}, \vec{q}) \Psi_n(\vec{p}-\vec{k},\vec{q}),
\label{FF2nc}
\eeq
and
\beq
{\mathcal F}_{nn}(k^2) = \int d^3p \ d^3q \ \Psi_c(\vec{p}, \vec{q}) \Psi_c(\vec{p}-\vec{k},\vec{q}).
\label{FF2nn}
\eeq

These relations can be simplified further by using the fact that at leading order only S-waves contribute. Consequently,
we project the three-body wave functions onto the S-wave, and then perform the angular integrations. In  our  
normalization, the wave functions then obey the relation:
\beq
\Psi_i(p,q) = 4\pi <\Psi_i(\vec{p},\vec{q})>,
\label{WFSwave}
\eeq
where $<...>$ denotes the angular average.
This relation is then substituted into the above form factor relations, and the trivial angular integrations 
can be performed.  Furthermore, the form factors will be normalized in the end such that ${\mathcal F}(k^2=0) = 1$, so 
any constant overall factor can be dropped. For the one-body form factors we have:
\beq
{\mathcal F}_i(k^2) = \int dp \ p^2 \int dq \ q^2 \int_{-1}^1 dx \ \Psi_i(p,q) \Psi_i(p,\sqrt{q^2+k^2-2qkx}),
\label{FFi}
\eeq
and for the two-body form factors we have:
\beq
{\mathcal F}_{nc}(k^2) = \int dp \ p^2 \int dq \ q^2 \int_{-1}^1 dx \ \Psi_n(p,q) \Psi_n(\sqrt{p^2+k^2-2pkx},q),
\label{FFnc}
\eeq
and
\beq
{\mathcal F}_{nn}(k^2) = \int dp \ p^2 \int dq \ q^2 \int_{-1}^1 dx \ \Psi_c(p,q) \Psi_c(\sqrt{p^2+k^2-2pkx},q).
\label{FFnn}
\eeq
The expressions relating $\Psi_n(p,q)$ and $\Psi_c(p,q)$ to the solutions $F_n(q)$ and $F_c(q)$
of Eqs.~(\ref{Fn}, \ref{Fc}) are given in Appendix \ref{sec:EigtoWF}. We are now in the position to calculate
these form factors for halo nuclei.

\begin{figure}[t]
	\centering
		\includegraphics*[width=10cm,angle=0]{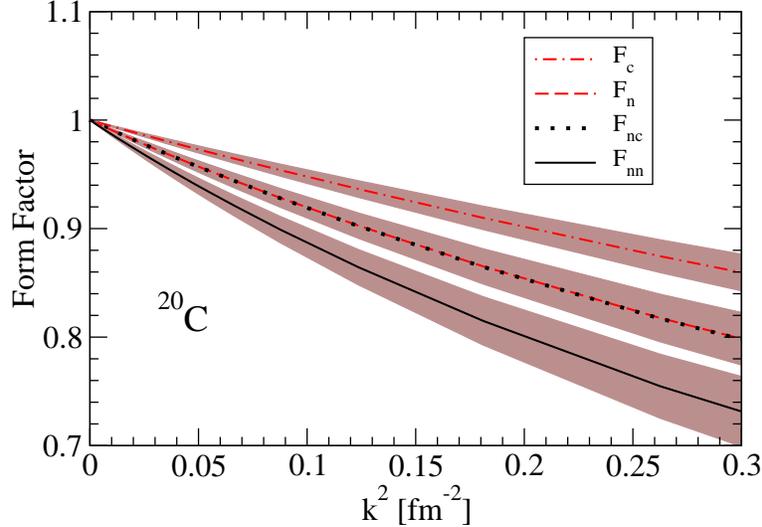}
	\caption{The various one- and two-body matter density form factors with leading order error bands 
	for the ground state of $^{20}$C in the low-energy 
	region: ${\mathcal F}_{nn}(k^2)$ black solid line; ${\mathcal F}_{nc}(k^2)$ black dotted line; 
	${\mathcal F}_{n}(k^2)$ lighter (red) dashed line; ${\mathcal F}_{c}(k^2)$ lighter (red) dot-dashed line.}
	\label{fig:FF20C}
\end{figure}

As a general example, we have plotted the form factors for the ground state of $^{20}$C, in the low-energy region in 
Fig.~\ref{fig:FF20C}, using a bound state energy of 3506.0 keV, a $n$-$n$ two-body virtual energy of 116.04 keV, and 
a $n$-$c$ bound state energy of 161.0 keV. 
The theory breaks down for momentum transfers of the order of the pion-mass 
squared ($k^2\approx 0.5$ fm$^{-2}$) where the 
one-pion exchange interaction cannot be approximated by short-range contact interactions anymore.
Here one can see that for low momentum transfer the one-body neutron, ${\mathcal F}_n(k^2)$, and 
the two-body core-neutron, ${\mathcal F}_{nc}(k^2)$, form factors lie nearly on top of each other. 
This is due to the fact that the 
core consists of $18$ nucleons and, therefore, the center of mass is very near the core. This fact is also 
seen in the shallow slope of the one-body core form factor, ${\mathcal F}_c(k^2)$. 

The theoretical error bands in the form factors are an estimate arising from the theoretical uncertainty of our 
two-body effective potential, Eq.~(\ref{SwaveVeff}). In this first order calculation, the uncertainty in 
the effective potential comes from the next term in the expansion, which is related to the effective range. 
Therefore, the theoretical uncertainty is $\approx r_0/a$. As discussed above, 
we use the inverse of the pion mass $m_{\pi} = 140$ MeV to approximate the 
effective range, $r_0 \approx 1/m_{\pi}$. With the form factors normalized to ${\mathcal F}(k^2=0) = 1$, 
and because $|E_{nc}| > |E_{nn}|$,
the theory error is estimated to be $\approx (1-{\mathcal F})\sqrt{2\mu_{nc} E_{nc}}/m_{\pi}$. 

\subsection{Mean square radii and geometry of 2$n$ halo nuclei}
\label{sec:Radii}

The mean square radii for our three-body bound states are calculated from the matter density form factors in the 
low momentum transfer region. The matter density form factor is defined as the Fourier transform of the 
matter density:
\beq
{\mathcal F}(k^2) = \int \rho(\vec{x}) e^{i \vec{k}\cdot \vec{x}} d^3x,
\label{FFdefined}
\eeq
with the normalization ${\mathcal F}(k^2=0) = 1$. 
In the low momentum transfer region, the exponential can be expanded, and assuming a spherically symmetric 
matter density, we see that the slope of the form factor determines the mean square 
radius $\left\langle r^2 \right\rangle$: 
\beq
{\mathcal F}(k^2) & = & \int \rho(\vec{x}) \left(1 + i \vec{k}\cdot \vec{x} - {(\vec{k}\cdot \vec{x})^2 \over 2} + \ldots \right) d^3x \nonumber\\
\nonumber\\
		& = & 1 - {1 \over 6} k^2 \left\langle r^2 \right\rangle + \ldots .
\label{FFexpand}
\eeq

\begin{figure}[t]
	\centering
		\includegraphics*[width=10cm,angle=0]{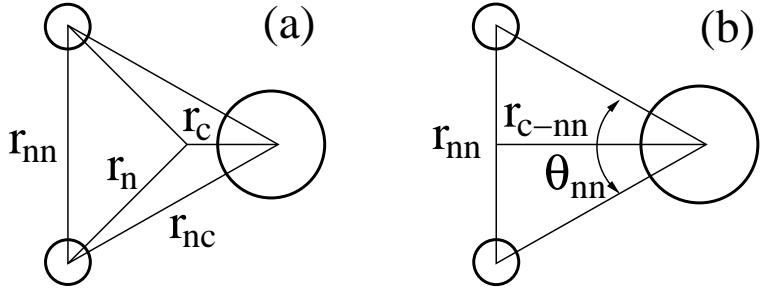}
	\caption{(a) The various radii of the three-body system. (b) Further geometry of the 2$n$ halo nucleus, 
	specifically looking at the two neutron opening angle $\theta_{nn}$.}
	\label{fig:Haloradii}
\end{figure}

Of course, the mean square radius acquired depends on the choice of one- or two-body form factor. 
Since $\vec{p}$ describes the relative momentum of the two particles in the two-body subsystem chosen, the slope 
of ${\mathcal F}_{ni}(k^2)$ will give the mean square distance between the two particles in the chosen two-body subsystem, 
either $\left\langle r_{nn}^2 \right\rangle$ or $\left\langle r_{nc}^2 \right\rangle$. Analogously, because $\vec{q}$ 
describes the momentum of the spectator particle relative to the center of mass of the two-body subsystem, the 
slope of ${\mathcal F}_{i}(k^2)$, will give the mean square distance of the spectator particle from the center of mass 
of the two-body subsystem, 
either $\left\langle r_{c-nn}^2 \right\rangle$ or $\left\langle r_{n-nc}^2 \right\rangle$. However, it is more 
useful to calculate the distance of the individual particles from the center of mass of the three-body bound 
state. If  $b_i$ is the slope of the one-body form factor ${\mathcal F}_{i}(k^2)$ at the limit $k^2 = 0$, 
the mean square radius of one 
of the bodies $i$ ($i = n,c$) from the three-body center of mass is given by:
\beq
\left\langle r_i^2 \right\rangle = -6b_i \left(1 - {m_i \over 2m_n + m_c} \right)^2,
\label{rfromb1}
\eeq
where $m_i$ is the mass of the desired particle $i$, and $m_n$ and $m_c$ are the neutron and core masses, 
respectively. The various radii of the three-body system are illustrated in Fig.~\ref{fig:Haloradii}(a).

We have extracted the radii by fitting a polynomial in $k^2$ to the form factor results  
for small $k^2$. We have used polynomials of varying degree up to 5th order in $k^2$ in order to verify 
the stability and convergence of the fit.
We have found a satisfactory stability in the slope when fitting to a polynomial 
to the fourth order in $k^2$, up to a value of $k^2$ at which the form factor has dropped less than 10 percent. 

As with the binding energies in the sequence of three-body Efimov states as discussed in Sec.~\ref{sec:Efimoveffect}, 
the mean square radii of these states also display universal properties that are insensitive to the details of the 
two-body potential at short distances. One such property is that for the 
resonant limit, $a \rightarrow \pm\infty$, at which there are infinitely many arbitrarily-shallow three-body 
bound states, the ratio of the radii of neighboring states approaches a universal factor as the threshold is approached. 
This universal scaling factor is exactly the inverse of the universal scaling factor found for the ratio of binding energies 
(see Eq.~(\ref{Eratio})):
\beq
\left\langle r^2 \right\rangle^{(n)}/\left\langle r^2 \right\rangle^{(n+1)} \longrightarrow ({\lambda_0})^{-2}, \ \textnormal{as} \ n \rightarrow +\infty \ \textnormal{with} \ a = \pm \infty.
\label{rratio}
\eeq
Therefore, we can construct a dimensionless quantity from the root of the product of the mean square radius and 
the three-body binding energy, $\sqrt{\left\langle r^2 \right\rangle B_3}$. 
The ratio of this quantity for neighboring states approaches unity in the resonant limit as the threshold is approached. 

The extracted radii for known halo nuclei are shown in Table~\ref{table:radii}. As input we have used the standard 
value of the $n$-$n$ scattering length, $a_{nn} = (-18.7 \pm 0.6)$ fm \cite{Gonzales-99}, to calculate 
the $n$-$n$ two-body virtual energy, $E_{nn} = 116.04$ keV, along with the experimental values 
of the $n$-$c$ two-body energies, $E_{nc}$, shown in the third column of Table~\ref{table:radii} 
(negative values correspond to virtual energies). 
As a three-body input, the cutoff is tuned to reproduce the experimental ground state binding energy, $B_3^{(0)}$, 
shown in the second column of Table~\ref{table:radii}. 
These experimental values for the two-body and three-body energies are taken from the most recent results of the 
"Nuclear Data Evaluation Project" of TUNL \cite{TUNL}, except where otherwise noted. 
In the last column, we have given the experimental values for the $n$-$n$ mean square radius, as given by 
Marqu\'es et al. \cite{Marques-00}. These experimental values for $\sqrt{\left\langle r_{nn}^2 \right\rangle}$ 
were found using the three-body correlation study in the dissociation of two neutrons in halo nuclei, 
along with the two neutron correlation function. However, the large uncertainty in these values is 
indicative of the need for more precise measurements of the mean square distances in 2$n$ halo nuclei. 

\begin{table}[!b]
	\begin{tabular}{||c|c|c||c|c|c|c||c||} \hline \hline
	Nucleus & $B_3$ [keV] & $E_{nc}$ [keV] & $\sqrt{\left\langle r_{nn}^2 \right\rangle}$ [fm] & 
        $\sqrt{\left\langle r_{nc}^2 \right\rangle}$ [fm] & $\sqrt{\left\langle r_{n}^2 \right\rangle}$ [fm] & 
        $\sqrt{\left\langle r_{c}^2 \right\rangle}$ [fm] & $\sqrt{\left\langle r_{nn}^2 \right\rangle}_{exp}$ [fm] \\ \hline 
	 	$^{11}$Li & 247 & -25 & 8.7$\pm$0.7 & 7.1$\pm$0.5 & 6.5$\pm$0.5 & 1.0$\pm$0.1 & \\
	 	& 247 & -800~\cite{Wilcox-75} & 6.8$\pm$1.8 & 5.9$\pm$1.5 & 5.3$\pm$1.4 & 0.9$\pm$0.2 & 6.6$\pm$1.5 \\
	 	& 320 & -800~\cite{Wilcox-75} & 6.2$\pm$1.6 & 5.3$\pm$1.4 & 4.8$\pm$1.3 & 0.8$\pm$0.2 & \\
	 	& 170 & -800~\cite{Wilcox-75} & 7.9$\pm$2.1 & 6.7$\pm$1.8 & 6.0$\pm$1.6 & 1.0$\pm$0.3 & \\ \hline
 	 $^{14}$Be & 1120 & -200 \cite{Thoennessen-00} & 4.1$\pm$0.5 & 3.5$\pm$0.5 & 3.2$\pm$0.4 & 0.40$\pm$0.05 & 5.4$\pm$1.0 \\ \hline 
	$^{12}$Be & 3673 & 503 & 3.0$\pm$0.6 & 2.5$\pm$0.5 & 2.3$\pm$0.5 & 0.32$\pm$0.07 & \\ \hline
 	 $^{18}$C & 4940 & 731 & 2.6$\pm$0.7 & 2.2$\pm$0.6 & 2.1$\pm$0.5 & 0.18$\pm$0.05 & \\ \hline
 	 $^{20}$C & 3506 & 161 & 2.8$\pm$0.3 & 2.4$\pm$0.3 & 2.3$\pm$0.3 & 0.19$\pm$0.02 & \\ 
  	& 3506 & 60 & 2.8$\pm$0.2 & 2.3$\pm$0.2 & 2.2$\pm$0.2 & 0.18$\pm$0.01 & \\ 
  	& 3506 & 0.0 & 2.7$\pm$0.2 & 2.2$\pm$0.2 & 2.1$\pm$0.2 & 0.18$\pm$0.01 & \\
  	$^{20}$C* & 65.0$\pm$6.8 & 60 & 42$\pm$3 & 38$\pm$3 & 41$\pm$3 & 2.2$\pm$0.2 & \\ 
 		$^{20}$C* & 1.02$\pm$0.08 & 0.0 & 130$\pm$10 & 97$\pm$7 & 93$\pm$7 & 6.9$\pm$0.5 & \\ \hline \hline
	\end{tabular}
\caption{\label{table:radii}
        Various mean square radii of different halo nuclei. The second two columns show the input values for 
        the three-body ground state energy and the two-body $n$-$c$ energy (negative values corresponding to 
        virtual energies), respectively, as given by \cite{TUNL}, except where otherwise noted.
        The experimental values for the $n$-$n$ root mean square radii, 
        shown in the last column, are taken from \cite{Marques-00}.
        The rows marked by $^{20}$C* show the results 
				for the excited Efimov state of $^{20}$C, with binding energy displayed in the second column, 
				which is found above the ground state ($B_3=3506$ keV).}
\end{table}

Our results agree overall with the study done by Yamashita et al. using a 
similar three-body model \cite{Yamashita:2004pv}. Our study expands on this previous work by showing the 
leading order theoretical uncertainty as well as the results for an excited Efimov state in the case of $^{20}$C. 

The leading order theoretical error is again estimated by the uncertainty of the two-body effective potential, 
Eq.~(\ref{SwaveVeff}), which is $\approx r_0/a$, where $r_0$ is the effective range of the interaction, and $a$ the 
scattering length. Using the inverse of the pion mass to estimate the effective range, $r_0 \approx 1/m_{\pi}$,
the uncertainty in the radii is then 
calculated from the greater of the error arising from the $n$-$n$ or $n$-$c$ interaction: 
$\sqrt{2\mu_{nc} E_{nc}}/m_{\pi}$ or $\sqrt{2\mu_{nn} E_{nn}}/m_{\pi}$.

\begin{figure}[t]
	\centering
		\includegraphics*[width=12cm,angle=0]{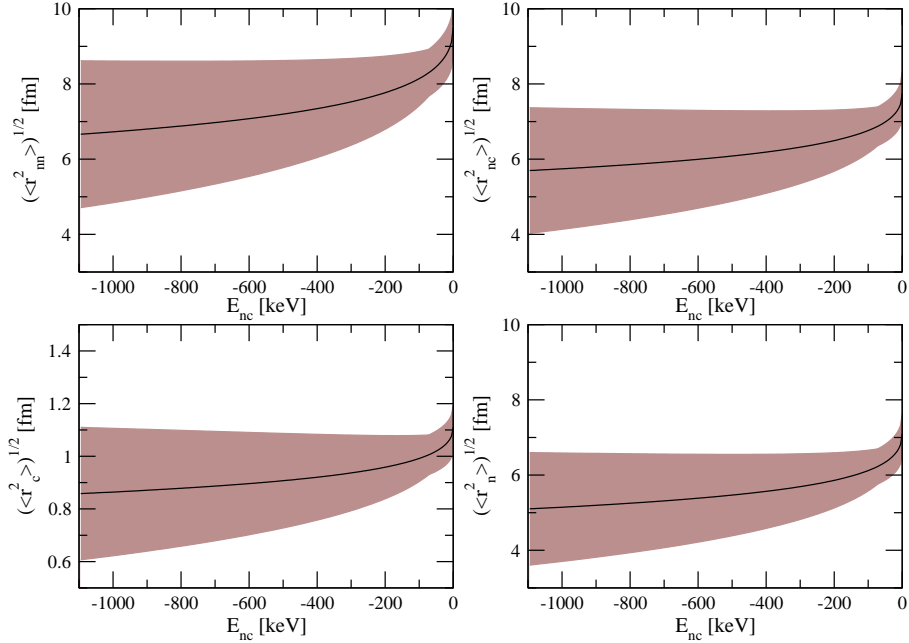}
	\caption{The various mean square radii for $^{11}$Li as a function of the $n$-$^9$Li two-body energy 
	(negative values correspond to the virtual state) with error bands from the theoretical uncertainty.
	As input, the $n$-$n$ two body energy $E_{nn} = -116.04$ keV, and the three-body binding energy $B_3^{(0)} = 247$ keV
	were used.}
	\label{fig:11LiRvsEnc}
\end{figure}

We will now discuss the results for the various known three-body halo nuclei as shown in Table~\ref{table:radii}:

For $^{11}$Li, there is a relatively 
large uncertainty in the experimental values of both the ground state energy, $B_3^{(0)} = (247\pm 80)$ keV \cite{TUNL}, 
and the $n$-$c$ virtual energy, with two competing values: $E_{nc} = (-25\pm 15)$ keV \cite{TUNL}, and $E_{nc}=(-800\pm 250)$ keV \cite{Wilcox-75}. 
For this reason it is advantageous to plot the various mean square radii over the full range of 
potential $E_{nc}$ values. The results, using the central value for the three-body binding energy as input, 
$B_3^{(0)} = 247$ keV, can be seen in Fig.~\ref{fig:11LiRvsEnc}, with error bands 
estimated from the theoretical uncertainty, as described above. 
These plots are a good general example of the relation between the mean square radii and the virtual two-body
$n$-$c$ energy for {\it Borromean} halo nuclei, where none of the two-body subsystems are bound.  As the $n$-$c$ 
virtual energy decreases in magnitude the three-body bound state increases slowly in size, with a more rapid 
increase in size as the energy approaches zero and crosses over into the {\it Samba} configuration, where 
the $n$-$c$ subsystem becomes bound. 

In Table~\ref{table:radii}, we have highlighted, using the central value of the three-body binding energy, 
the central values of the competing $n$-$c$ energies.
While the two-body virtual energy  reported in \cite{Wilcox-75}, $E_{nc} = -800$ keV leads to 
$\sqrt{\left\langle r_{nn}^2 \right\rangle} = (6.8\pm 1.8)$ fm in close agreement with the experimental 
result $\sqrt{\left\langle r_{nn}^2 \right\rangle}_{exp} = (6.6\pm 1.5)$ fm
a definite conclusion can not be reached due to the large error bars of the radii.
We have also listed the upper and lower limits of the three-body binding energy, $B_3^{(0)} = 170$ and 320 keV, 
along with our preferred value of $E_{nc} = -800$ keV, which shows a more general result: for 
halo nuclei, the larger the three-body binding energy, the smaller the mean square radii. In terms of the 
plots in Fig.~\ref{fig:11LiRvsEnc}, using a larger (smaller) value for $B_3^{(0)}$ as input would shift 
the curve down (up) in each plot. Due to the large uncertainties in both the theoretical and experimental values 
for $^{11}$Li, there exists a large range of $E_{nc}$ values which 
would produce a $\sqrt{\left\langle r_{nn}^2 \right\rangle}$ value 
in agreement with the experimental value of 
Marqu\'es et al.~\cite{Marques-00}.

As another example of a {\it Borromean} halo nucleus, we see that the calculated result for the $n$-$n$ mean 
square radius of $^{14}$Be, $\sqrt{\left\langle r_{nn}^2 \right\rangle} = (4.1\pm 0.5)$ fm is smaller than the 
experimental value $\sqrt{\left\langle r_{nn}^2 \right\rangle}_{exp} = (5.4\pm 1.0)$ fm~\cite{Marques-00}, 
but still within one error bar. 
Here we have used the central value of the two-body $n$-$c$ virtual energy 
as reported by \cite{Thoennessen-00}, $E_{nc} = -200$ keV. In using the resonant limit, $E_{nc} = 0.0$ keV, we 
see that the largest theoretical value for $\sqrt{\left\langle r_{nn}^2 \right\rangle} = (4.6\pm 0.3)$ fm, which allows an 
unbound $n$-$c$ two-body subsystem, is still smaller than the experimental value. 
Another reported value for the two-body virtual energy, $E_{nc} = (-1900\pm 500)$ keV
\cite{TUNL}, would produce even smaller mean square radii, even farther away from the experimental value. 

We now turn our attention to the so called {\it Samba} halo nuclei in which the $n$-$c$ subsystem is bound. 
As examples, we have listed the results for $^{12}$Be and $^{18}$C, using the central values of the experimental 
energies \cite{TUNL} in Table~\ref{table:radii}. 

\begin{figure}[t]
	\centering
		\includegraphics*[width=12cm,angle=0]{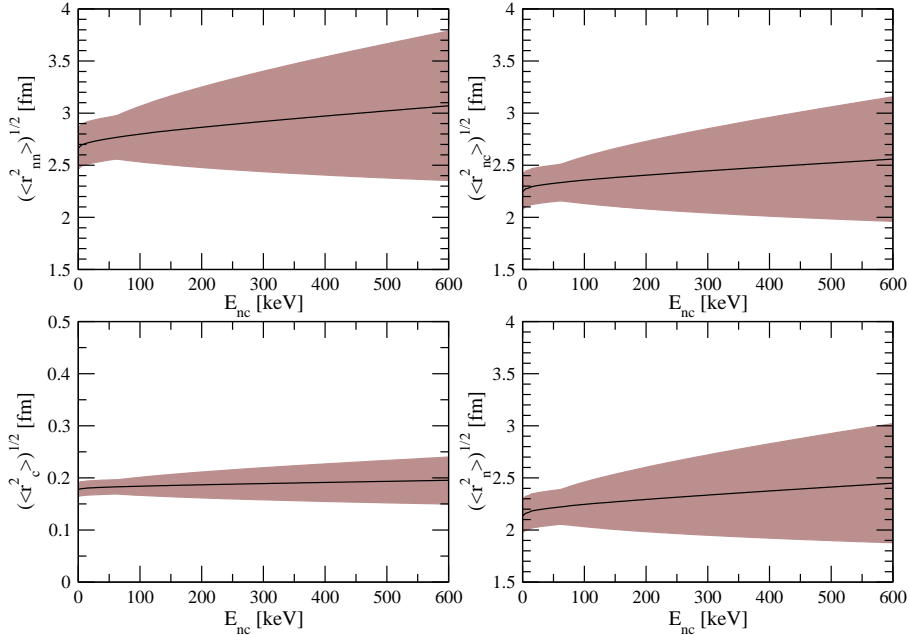}
	\caption{The various mean square radii for $^{20}$C as a function of the $n$-$^{18}$C two-body energy 
	with error bands from the theoretical uncertainty.
	As input, the $n$-$n$ two body energy $E_{nn} = -116.04$ keV, and the three-body binding energy $B_3^{(0)} = 3506$ keV
	were used.}
	\label{fig:20CRvsEnc0}
\end{figure}

Of greater interest are the results from the case of $^{20}$C, as the large uncertainty in the $n$-$c$ energy,
with two competing values, $E_{nc} = (162 \pm 112)$ keV \cite{TUNL}, and $E_{nc} = 530$ keV \cite{Nakamura-99}, suggests that we look at the 
mean square radii over a range of $E_{nc}$ values.  The results, using the central value for the 
three-body binding energy as input, $B_3^{(0)} = 3506$ keV, can be seen in Fig.~\ref{fig:20CRvsEnc0}, with error bands 
estimated from the theoretical uncertainty, as described above. 
These plots are a good general example of the relation between the mean square radii and the two-body
$n$-$c$ binding energy for {\it Samba} halo nuclei.  As the $n$-$c$ binding
energy decreases in magnitude the three-body bound state decreases slowly in size, with a slightly more rapid 
decrease in size as the energy approaches zero and crosses over into the {\it Borromean} configuration, where 
the $n$-$c$ subsystem becomes unbound. This suggests that as the two-body $n$-$c$ state is more weakly bound, 
the particles must be closer together in order for the three-body state to be bound with the same energy. 

\begin{figure}[t]
	\centering
		\includegraphics*[width=12cm,angle=0]{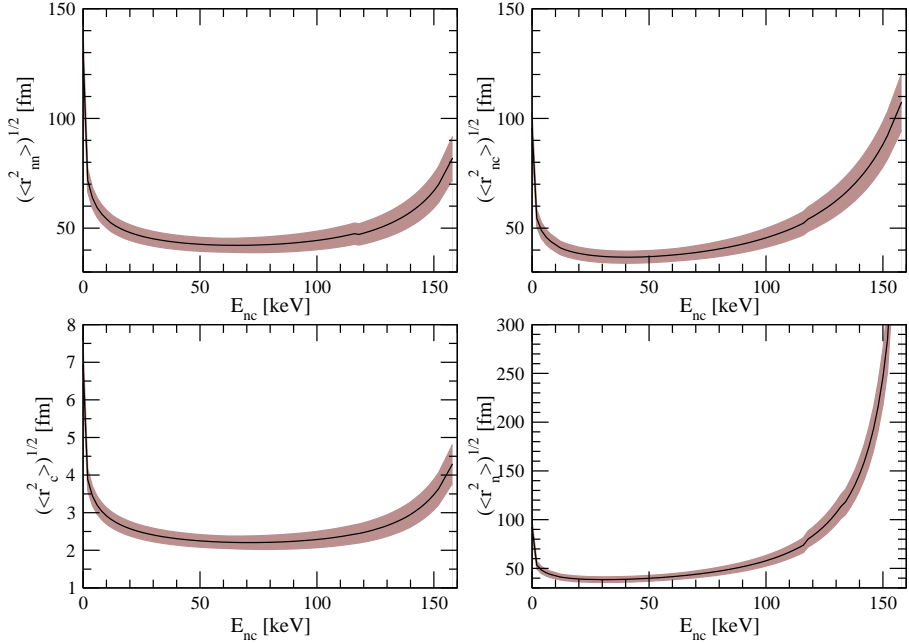}
	\caption{The various mean square radii for the Efimov excited state of $^{20}$C as a function of the 
	$n$-$^{18}$C two-body energy with error bands from the theoretical uncertainty.
	As input, the $n$-$n$ two body energy $E_{nn} = -116.04$ keV, and the three-body binding energy $B_3^{(0)} = 3506$ keV
	were used.}
	\label{fig:20CRvsEnc1}
\end{figure}

As was shown in Sec.~\ref{sec:Carbon20}, there possibly exists one Efimov excited state in $^{20}$C for $E_{nc} < 165$ keV. 
The mean square radii for this excited state were calculated over a range of $E_{nc}$ values and plotted, with leading 
order error bands, in Fig.~\ref{fig:20CRvsEnc1}. 
Here we see the interesting phenomenon that although the radii remain relatively constant in the middle of the 
range of $n$-$c$ energies which allow the excited state, as the endpoints are approached, the radii begin to 
increase rapidly, and then to diverge, both when $E_{nc} \rightarrow 165$ and $\rightarrow 0.0$ keV. It is at these points that 
the $^{20}$C system moves outside the boundary curve depicted in Fig.~\ref{fig:KnKaerrors}, the Efimov excited 
state is destroyed, and the three particles consequently fly apart. 

In Table~\ref{table:radii}, we have highlighted the result using the central value of the accepted 
$n$-$c$ two-body binding energy, as well as two values which lead to an Efimov excited state, 
including the resonant limit $E_{nc} = 0.0$ keV. The rows marked by $^{20}$C$^*$ represent the results 
of the Efimov excited state. The three-body binding energy 
of this excited state is listed in the second column with leading order threoretical uncertainty calculated as 
described in Sec.~\ref{sec:Carbon20}. 

Next we looked at a further geometrical property of 2$n$ halo nuclei, specifically the two neutron opening angle. 
As defined in Fig.~\ref{fig:Haloradii}(b), it is straight forward to calculate $\theta_{nn}$ using the pair of mean square radii 
found from the three-body wave function $\Psi_c$, $\sqrt{\left\langle r_{nn}^2 \right\rangle}$ and 
$\sqrt{\left\langle r_{c-nn}^2 \right\rangle}$:
\beq
\tan \left({\theta_{nn} \over 2} \right) = {{1 \over 2}\sqrt{\left\langle r_{nn}^2 \right\rangle} \over \sqrt{\left\langle r_{c-nn}^2 \right\rangle}}.
\label{nnangle}
\eeq
Our results for known halo nuclei are shown in Table~\ref{table:angles} using the central values of the experimental 
two-body and three-body energies as inputs (compare with Table~\ref{table:radii}). We show the results for the competing 
values of the $n$-$c$ virtual energy for the case of $^{11}$Li.  For the case of $^{20}$C we also show the result 
with $E_{nc} = 0.0$ keV, along with its corresponding Efimov excited state. The opening angle has been calculated 
from experimental data in two recent works by Bertulani et al.~\cite{Bertulani-07}, 
and Hagino et al.~\cite{Hagino-07}, and their results are shown in the last two columns, respectively. 
The study by Bertulani et al. uses the experimental values of $\sqrt{\left\langle r_{nn}^2 \right\rangle}$ 
found by Marqu\'es et al.~\cite{Marques-00} seen in the last column of Table~\ref{table:radii}, along with two 
different determinations of $\sqrt{\left\langle r_{c-nn}^2 \right\rangle}$ 
(see \cite{Bertulani-07} and references within): 
using laser spectroscopy data (results displayed in first row for $^{11}$Li and $^{14}$Be), 
and also using the $B(E1)$ strength (second row for $^{11}$Li). However the result 
for $^{14}$Be also uses a theoretical calculation for one of the inputs, rather than being a pure experimental 
result. On the other hand, the study by Hagino et al.~uses the experimental values of $B(E1)$ to calculate 
$\sqrt{\left\langle r_{c-nn}^2 \right\rangle}$, along with two different determinations of 
$\sqrt{\left\langle r_{nn}^2 \right\rangle}$ (see \cite{Hagino-07} and references within): 
using experimental values of the nuclear matter radii (first row for $^{11}$Li), and using the 
results of Marqu\'es et al.~\cite{Marques-00} (second row for $^{11}$Li).  Also, 
results found using a three-body model with density dependant two-body contact interactions were found in 
\cite{Hagino-07}, and are displayed in the third row for $^{11}$Li. As would be expected, our results 
agree very well with the results from the three-body theoretical model used in \cite{Hagino-07}. Our result for 
$^{11}$Li using our preferred choice of $E_{nc} = -800$ keV, also agrees very well with the experimental results 
obtained using the value for $\sqrt{\left\langle r_{nn}^2 \right\rangle}_{exp}$ from Marqu\'es et al., which would 
be expected as our $\sqrt{\left\langle r_{nn}^2 \right\rangle}$ value also agrees with this experimental value. 
Overall there is a good agreement between our calculated results and the results of \cite{Bertulani-07} and 
\cite{Hagino-07}, as all $\theta_{nn}$ values lie within one error bar of each other. However, the size of these 
error bars suggest that further study should be done to improve both the experimental and theoretical results. 

\begin{table}[!t]
	\begin{tabular}{||c|c|c||c||c|c||} \hline \hline
		Nucleus & $\sqrt{\left\langle r_{nn}^2 \right\rangle}$ [fm] & $\sqrt{\left\langle r_{c-nn}^2 \right\rangle}$ [fm] &	
			$\theta_{nn}$ & $\theta_{nn}$ \cite{Bertulani-07} & $\theta_{nn}$ \cite{Hagino-07} \\ \hline
		$^{11}$Li & 8.7$\pm$0.7 & 5.5$\pm$0.4 & $77^{\circ +8}_{-9}$ & $58^{\circ +10}_{-14}$ & $56.2^{\circ +17.8}_{-21.3}$ \\
			& 6.8$\pm$1.8 & 5.0$\pm$1.3 & $68^{\circ +31}_{-25}$ & $66^{\circ +22}_{-18}$ & $65.2^{\circ +11.4}_{-13.0}$ \\
		        &	&	& 	& 	& [65.29] \\ \hline
		$^{14}$Be & 4.1$\pm$0.5 & 2.8$\pm$0.4 & $72^{\circ +16}_{-13}$ & $64^{\circ +9}_{-10}$ & \\ \hline
		$^{12}$Be & 3.0$\pm$0.6 & 1.9$\pm$0.4 & $77^{\circ +23}_{-22}$ &	&	\\ \hline
		$^{18}$C  & 2.6$\pm$0.7 & 1.6$\pm$0.4 & $78^{\circ +30}_{-27}$ &	&	\\ \hline
		$^{20}$C  & 2.8$\pm$0.3 & 2.0$\pm$0.2 & $70^{\circ +11}_{-11}$ &	&	\\
 			& 2.7$\pm$0.2 & 1.8$\pm$0.1 & $74^{\circ +7}_{-7}$ &	&	\\
 		$^{20}$C* & 130$\pm$10	& 69$\pm$5		& $87^{\circ +8}_{-9}$ &	&	\\ \hline \hline
 		 \end{tabular}
\caption{\label{table:angles}
        Various two neutron opening angles of different 2$n$ halo nuclei calculated from the results for 
        the mean square radii shown.
        Compared with results of \cite{Bertulani-07} and \cite{Hagino-07} shown in the last two columns, respectively.}
\end{table}

\section{Conclusion}
\label{sec:Conclusion}

In this paper, we have investigated universal aspects of three-body halo nuclei within an effective quantum mechanics approach.
Assuming that the halo nuclei have resonant S-wave interactions between the neutron and the core, the effective potential at leading
order reduces to a separable S-wave potential.
The corrections at next-to- and next-to-next-to-leading order in the 
expansion in $M_{low}/M_{high}$ are determined by the S-wave effective
ranges \cite{Hammer:2001gh,Platter:2006ev}. Corrections from P-wave
interactions appear at even higher orders \cite{Bedaque:2002mn}.
An important improvement compared to previous calculations is the inclusion of error bands based on
the omitted higher order terms in the effective theory.

We have calculated the parametric region within which at least one excited Efimov state will occur for different values of the core mass $A$.
The boundary of this region is given by a curve in the plane described by 
the root of the ratio of the two-body bound(virtual) state energies to 
the ground state energy~\cite{Amorim-96}. We have calculated the
boundary of this region for various values of the core mass $A$
and provided error bands for the boundary curves.
From the current experimental data, we conclude that none of the known halo 
nuclei is likely to have an excited Efimov state.
One possible exception is $^{20}$C which could have one excited state with a binding energy of less than 7 keV. 

Next, we have studied the structure of known 2$n$ halo nuclei, calculating the one- and two-body matter density form 
factors.  From these form factors we were able to extract the mean square distances between the two particles in the chosen 
two-body subsystem, as well as the mean square distance of the spectator particle from the center of mass. We found that 
our results for the $n$-$n$ mean square radius agree well with the experimental data for the {\it Borromean} halo nuclei
$^{11}$Li and $^{14}$Be \cite{Marques-00}. We have explicitly not studied the case of $^6$He, which is dominated by a P-wave resonance in the 
$n$-$c$ interaction (``$^5$He'') and requires a different counting scheme.
While various schemes to treat such P-wave resonances in Effective Field Theory have been developed
\cite{Bertulani-02,BHvK2}, their application to three-body systems remains to 
be worked out.
To the expected accuracy, our effective theory gives a good description  of the studied halo nuclei. Using our results for the mean square distances, 
we have also calculated the two neutron opening angle, and found a good general agreement with the recent results of \cite{Bertulani-07} and 
\cite{Hagino-07}. 

Throughout, this work,  we have estimated the theoretical error of the leading order effective potential, Eq.~(\ref{SwaveVeff}). This 
uncertainty was quantified in our results through error bands. A future study could systematically improve the theoretical error through the 
inclusion of a momentum dependent next-to-leading order term in the effective potential which can be matched to the effective range of the 
interactions. 

Another interesting application of this effective theory will be the 
study of Coulomb excitation data from existing and future facilities 
with exotic beams (such as FAIR and FRIB).
In these experiments a nuclear beam scatters 
off the Coulomb field of a heavy
nucleus. Such processes can populate excited states of the projectile
which subsequently decay, leading to its ``Coulomb dissociation''
\cite{Bert88}. Effective theories offer a systematic
framework for a full quantum-mechanical treatment of these reactions.
In summary, with new improved experimental data for these weakly bound 
nuclei, much more knowledge can be obtained about the structure of these 
interesting systems as well as discovering 
whether they show universal behavior and excited Efimov states. 

\begin{acknowledgments}

We thank Lucas Platter and Andreas Nogga for helpful discussions
and Ulf Mei\ss ner for comments on the manucript. 
This research was supported by the
BMBF under contract number 06BN411.

\end{acknowledgments}

\appendix

\section{Reconstruction of the wave function}
\label{sec:EigtoWF}

In this section we give the expressions for the S-wave part of the full wave functions $\Psi_n(p,q)$ and $\Psi_c(p,q)$ where 
the index $i=n,c$ labels the chosen spectator particle. 
In the wave functions, the $p$ Jacobi momentum describes the relative momentum between 
the two particles in the chosen two-body subsystem, while $q$ describes the momentum of the spectator 
particle relative to the center of mass of the two-body subsystem.

For the neutron as the spectator particle, we find:
\beq
\Psi_n(p,q) 	& = & G^n_0(p,q;B_3) g(p) t_n(q;B_3) F_n(q) \nonumber\\
\nonumber\\
		& & + {1 \over 2} \int_{-1}^{1} dx \ G^n_0(\tilde{\pi}_{nn}, \tilde{\pi}'_{nn}; B_3) g(\tilde{\pi}_{nn}) 
t_n(\tilde{\pi}'_{nn};B_3) F_n(\tilde{\pi}'_{nn}) \nonumber\\
\nonumber\\
		& & + {1 \over 2} \int_{-1}^{1} dx \ G^c_0(\tilde{\pi}_{nc}, \tilde{\pi}'_{nc}; B_3) g(\tilde{\pi}_{nc}) 
t_c(\tilde{\pi}'_{nc};B_3) F_c(\tilde{\pi}'_{nc}),
\label{PsinF}
\eeq
where the regulator function $g$ and the T-matrices $t_n$ and $t_c$ are given in Eqs.~(\ref{reg}, \ref{tn}, \ref{tc}).
The expressions for the propagators $G_0^n$ and $G_0^c$ can be found in Eqs.~(\ref{propn}) and (\ref{propc}).
One can show through a simple substitution that $G^n_0(\tilde{\pi}_{nn}, \tilde{\pi}'_{nn};B_3) = G^n_0(p,q;B_3)$, and
$G^c_0(\tilde{\pi}_{nc}, \tilde{\pi}'_{nc};B_3) = G^n_0(p,q;B_3)$, and the propagators simplify.
Finally, the shifted momenta are:
\bqa
\tilde{\pi}_{nn} &\equiv& \tilde{\pi}_{nn}(p,q) = \sqrt{{1\over (A+1)^2}\, p^2 + {A^2 (A+2)^2 \over (A+1)^4}\, 
q^2 + {A (A+2) \over (A+1)^3}\, 2pqx},
\\
\tilde{\pi}'_{nn} &\equiv& \tilde{\pi}'_{nn}(p,q) = \sqrt{p^2 + {1 \over (A+1)^2}\, q^2 - {1\over A+1}\,2 pqx},
\\
\tilde{\pi}_{nc} &\equiv& \tilde{\pi}_{nc}(p,q) = \sqrt{{1 \over 4}\, p^2 + {(A+2)^2 \over 4 (A+1)^2} \,q^2 
+ {A+2 \over 2(A+1)} pqx},
\\
\tilde{\pi}'_{nc} &\equiv& \tilde{\pi}'_{nc}(p,q) = \sqrt{p^2 + {A^2 \over (A+1)^2}\, q^2 - {A\over A+1} 2pqx}.
\eqa

If the core is the spectator particle, we find:
\beq
\Psi_c(p,q) 	& = & \int_{-1}^{1} dx \ G^n_0(\tilde{\pi}_{cn}, \tilde{\pi}'_{cn}; B_3) g(\tilde{\pi}_{cn}) t_n(\tilde{\pi}'_{cn};B_3) 
F_n(\tilde{\pi}'_{cn}) \nonumber\\
\nonumber\\
		& & + G^c_0(p,q;B_3) g(p) t_c(q;B_3) F_c(q).
\label{PsicF}
\eeq
Again, one can show through a simple substitution that $G^n_0(\tilde{\pi}_{cn}, \tilde{\pi}'_{cn};B_3) = G^c_0(p,q;B_3)$.
For this case, the shifted momenta are given by:
\bqa
\tilde{\pi}_{cn} &\equiv& \tilde{\pi}_{cn}(p,q) = \sqrt{ {A^2\over (A+1)^2}\, p^2 + { (A+2)^2 \over 4 (A+1)^2} \, q^2 
+ {A (A+2) \over 2 (A+1)^2}\, 2pqx},
\\
\tilde{\pi}'_{cn} &\equiv& \tilde{\pi}'_{cn}(p,q) = \sqrt{p^2 + {1 \over 4} q^2 - pqx}.
\eqa



\begin{thebibliography}{99}

\bibitem{Braaten-05}
  E.~Braaten and H.-W.~Hammer,
  Phys.\ Rept.\  {\bf 428} (2006) 259
  [arXiv:cond-mat/0410417].

\bibitem{Bedaque:1998kg}
P.F.~Bedaque, H.-W.~Hammer, and U.~van Kolck,
Phys.\ Rev.\ Lett.\  {\bf 82} (1999) 463
[arXiv:nucl-th/9809025];
Nucl.\ Phys.\ A {\bf 646} (1999) 444
[arXiv:nucl-th/9811046].

\bibitem{Efimov-70}
V. Efimov, 
Phy. Lett. {\bf 33B} (1970) 563.

\bibitem{Bedaque:2002mn}
P.F.~Bedaque and U.~van Kolck,
Ann.\ Rev.\ Nucl.\ Part.\ Sci.\  {\bf 52} (2002) 339
[arXiv:nucl-th/0203055].

\bibitem{Bedaque:2002yg}
  P.F.~Bedaque, G.~Rupak, H.W.~Griesshammer, and H.-W.~Hammer,
  Nucl.\ Phys.\ A {\bf 714} (2003) 589
  [arXiv:nucl-th/0207034].


\bibitem{Platter:2004zs}
  L.~Platter, H.-W.~Hammer, and U.-G.~Mei\ss ner,
  Phys.\ Lett.\ B {\bf 607} (2005) 254
  [arXiv:nucl-th/0409040].
  

\bibitem{Riisager-94}
K. Riisager,
Rev. Mod. Phys. {\bf 66} (1994) 1105.

\bibitem{Zhukov-93}
M.V. Zhukov, B.V. Danilin, D.V. Fedorov, J.M. Bang, I.J. Thompson, 
and J.S. Vaagen, 
Phys. Rep. {\bf 231} (1993) 151.

\bibitem{Hansen-95}
P.G. Hansen, A.S. Jensen, and B. Jonson, Ann.\ Rev.\ Nucl.\ Part. \ Sci.\
{\bf 45} (1995) 591.

\bibitem{Taniha-96}
I. Tanihata, J.\ Phys.\ G {\bf 22} (1996) 157.

\bibitem{Jensen-04}
A.S. Jensen, K. Riisager, D.V. Fedorov, and E. Garrido, 
Rev. Mod. Phys. {\bf 76} (2004) 215.

\bibitem{Bertulani-02}
  C.A.~Bertulani, H.-W.~Hammer and U.~Van Kolck,
  Nucl.\ Phys.\  A {\bf 712} (2002) 37
  [arXiv:nucl-th/0205063].

\bibitem{BHvK2}
P.F.~Bedaque, H.-W.~Hammer and U.~van Kolck,
  Phys.\ Lett.\  B {\bf 569} (2003) 159
  [arXiv:nucl-th/0304007].

\bibitem{BRvK}
C.A. Bertulani, R. Higa, and U. van Kolck, in progress. 

\bibitem{Higa:2008dn}
  R.~Higa, H.-W.~Hammer and U.~van Kolck,
  Nucl.\ Phys.\ A (in press)
  [arXiv:0802.3426 [nucl-th]].

  \bibitem{Hammer:2008ra}
  H.-W.~Hammer and R.~Higa,
  Eur.\ Phys.\ J.\ A (in press) [arXiv:0804.4643 [nucl-th]].

\bibitem{Kraemer-06}
T.~Kraemer et al., 
Nature {\bf 440} (2006) 315 [arXiv:cond-mat/0512394v2].


\bibitem{Fedorov-94}
  D.V.~Fedorov, A.S.~Jensen and K.~Riisager,
  Phys.\ Rev.\ Lett.\  {\bf 73} (1994) 2817
  [arXiv:nucl-th/9409018].
  
\bibitem{Amorim-96}
  A.E.A.~Amorim, T.~Frederico and L.~Tomio,
  Phys. Rev. C {\bf 56} (1997) R2378
  [arXiv:nucl-th/9708023].

\bibitem{Mazumdar-00}
I. Mazumdar, V. Arora, and V. S. Bhasin, 
Phys. Rev. C {\bf 61} (2000) R051303.

\bibitem{Platter:2004qn}
  L.~Platter, H.-W.~Hammer and U.-G.~Mei\ss ner,
  Phys.\ Rev.\  A {\bf 70} (2004) 052101
  [arXiv:cond-mat/0404313].

\bibitem{Lucas-04}
L.~Platter, H.-W.~Hammer and U.-G.~Mei\ss ner,
  Few Body Syst.\  {\bf 35} (2004) 169
  [arXiv:cond-mat/0405660].

\bibitem{Kharchenko-73}
V.F. Kharchenko, 
Sov. J. Nucl. Phys. {\bf 16} (1973) 173 [Yad. Fiz. {\bf 16} (1972) 310].

\bibitem{Efimov-90}
V. Efimov, Comments Nucl. Part. Phys. {\bf 19} (1990) 271.

\bibitem{Thomas-35}
L.H. Thomas, 
Phys. Rev. {\bf 47} (1935) 903.

\bibitem{Adhikari-88}
S.K. Adhikari, A. Delfino, T. Frederico, I.D. Goldman, and L. Tomio, 
Phys. Rev. A {\bf 37} (1988) 3666.

\bibitem{Robicheaux-99}
F. Robicheaux, Phys. Rev. A {\bf 60} (1999) 1706.

\bibitem{Yamashita-Samba}
M.T. Yamashita, T. Frederico, and M.S. Hussein, 
  Mod.\ Phys.\ Lett.\  A {\bf 21} (2006) 1749
  [arXiv:nucl-th/0501052].

\bibitem{Tiesinga-93}
E. Tiesinga, B.J. Verhaar, and H.T.C. Stoof, 
Phys. Rev. A {\bf 47} (1993) 4114.

\bibitem{TUNL} TUNL nuclear data evaluation project. WWW: http://www.tunl.duke.edu/NuclData/.

\bibitem{Gonzales-99}
D.E. Gonzales Trotter {\it et al.}, Phys. Rev. Lett. {\bf 83} (1999) 3788.

\bibitem{Audi-95}
G. Audi and A.H. Wapstra, Nucl. Phys. {\bf A595} (1995) 409.

\bibitem{Yamashita-07}
  M.T.~Yamashita, T.~Frederico and L.~Tomio,
  Phys.\ Lett.\ B {\bf 660} (2008) 339
  [arXiv:0704.1461 [nucl-th]].

\bibitem{Arora-04}
V. Arora, I. Mazumdar, and V. S. Bhasin, 
Phys. Rev. C {\bf 69} (2004) R061301.


\bibitem{Mazumdar-06}
  I.~Mazumdar, A.R.P.~Rau and V.~S.~Bhasin,
  Phys.\ Rev.\ Lett.\  {\bf 97} (2006) 062503
  [arXiv:quant-ph/0607193].

\bibitem{Marques-00}
  F.M.~Marqu\'{e}s et al.,
  Phys.\ Lett.\  B {\bf 476} (2000) 219;
  Phys.\ Rev.\  C {\bf 64} (2001) 061301.
  
\bibitem{Yamashita:2004pv}
  M.T.~Yamashita, L.~Tomio and T.~Frederico,
  Nucl.\ Phys.\  A {\bf 735} (2004) 40
  [arXiv:nucl-th/0401063].
  
\bibitem{Wilcox-75}
  K.H.~Wilcox et al.,
  Phys.\ Lett.\ {\bf 59B} (1975) 142.
  
\bibitem{Thoennessen-00}
  M.~Thoennessen, S.~Yokoyama, P.~G.~Hansen,
  Phys.\ Rev.\ C {\bf 63} (2000) 014308.
  
\bibitem{Nakamura-99}
  T.~Nakamura et al., 
  Phys.\ Rev.\ Lett.\ {\bf 83} (1999) 1112.
  
\bibitem{Bertulani-07}
  C.A.~Bertulani and M.S.~Hussein,
  Phys.\ Rev.\  C {\bf 76} (2007) 051602
  [arXiv:0705.3998 [nucl-th]].
  
\bibitem{Hagino-07}
  K.~Hagino and H.~Sagawa,
  Phys.\ Rev.\  C {\bf 76} (2007) 047302
  [arXiv:0708.1543 [nucl-th]].

\bibitem{Hammer:2001gh}
  H.-W.~Hammer and T.~Mehen,
  Phys.\ Lett.\  B {\bf 516} (2001) 353.

\bibitem{Platter:2006ev}
  L.~Platter and D.R.~Phillips,
Few Body Syst.\  {\bf  40} (2006) 35.

\bibitem{Bert88}
C.A.~Bertulani and G.~Baur, Phys.\ Rep.\ {\bf 163} (1988) 299.

\end{thebibliography}
\end{document}